\documentclass[a4paper]{jpconf}
\usepackage{graphicx}
\usepackage{cite}

\begin{document}

\title{Low-temperature thermodynamics of one class of flat-band models}

\author{O Derzhko$^1$, J Richter$^2$ and A Honecker$^3$}

\address{$^1$Institute for Condensed Matter Physics, National Academy of Sciences of Ukraine,\\
             1 Svientsitskii Street, L'viv-11, 79011, Ukraine}
\address{$^2$Institut f\"ur Theoretische Physik, Universit\"at Magdeburg,\\
             P.O. Box 4120, 39016 Magdeburg, Germany}
\address{$^3$Institut f\"ur Theoretische Physik, Universit\"at G\"ottingen, 37077 G\"ottingen, Germany}

\ead{derzhko@icmp.lviv.ua}

\begin{abstract}
We consider the antiferromagnetic Heisenberg model and the repulsive Hubbard model
for a class of frustrated lattices with a completely dispersionless (flat) lowest one-particle
(either one-magnon or one-electron)
band.
We construct exact many-particle ground states for a wide range of particle densities,
calculate their degeneracy,
and, as a result, obtain closed-form expressions for the low-temperature thermodynamic quantities
around a particular value of the magnetic field $h_{\rm{sat}}$ or the chemical potential $\mu_0$.
We confirm our analytic findings by numerical data for finite lattices.
\end{abstract}

\section{Introduction}

Strongly correlated systems on geometrically frustrated lattices
represent a playground to study many collective quantum phenomena.
In this paper,
we consider a particular class of geometrically frustrated lattices,
namely lattices which support flat (i.e.\ completely dispersionless) one-particle bands.
The antiferromagnetic Heisenberg model on such lattices was examined in Refs.~\cite{prl,zhito:2004,dr:2004,ltp},
although in the context of flat-band ferromagnetism
some of these lattices were discussed even earlier \cite{mielke,tasaki:1998}, see  also Ref.~\cite{dhr}.
The flat one-particle band leads to the possibility
to localize the corresponding one-particle states within a finite region of a lattice.
Considering further many-particle states one may expect
that the states with localized particles which are spatially separated from each other
are also the eigenstates of the Hamiltonian with interaction.
Under certain conditions a manifold of localized states may constitute
a highly degenerate ground state of the interacting many-particle system
and as a result the localized states may dominate the low-temperature thermodynamics.

\begin{figure}
\begin{center}
\includegraphics[width=17pc]{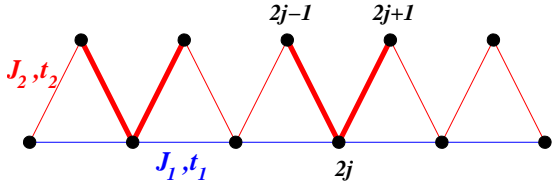}
\hspace{2pc}
\begin{minipage}[b]{17pc}
\caption{\label{fig1}
The sawtooth chain.
The exchange (hopping) integrals are $J_1$ ($t_1$) along the base line and $J_2$ ($t_2$) along the zigzag line.
The bold valleys show the area occupied by localized magnons (electrons).}
\end{minipage}
\end{center}
\end{figure}

In the present paper we compare and contrast
the consideration of the localized-states effect for the low-temperature thermodynamics for two models,
the spin-1/2 antiferromagnetic Heisenberg model
and
the one-orbital repulsive Hubbard model.
For concreteness we focus on the sawtooth-chain ($\Delta$-chain) lattice shown in Fig.~\ref{fig1}
(for other lattices see Ref.~\cite{ltp}).
More specifically, we deal with the antiferromagnetic Heisenberg Hamiltonian
\begin{eqnarray}
\label{heis_ham}
H_{{\rm{Heis}}}
=\sum_{\langle i, j\rangle}J_{i,j}
\left[
\frac{1}{2}\left(s_i^+s_j^-+s_i^-s_j^+\right)+s_i^zs_j^z
\right]
-h\sum_is_i^z
\end{eqnarray}
with the nearest-neighbor exchange integrals $J_{i,j}>0$,
and the Hubbard Hamiltonian
\pagebreak
\begin{eqnarray}
\label{hubb_ham}
H_{{\rm{Hub}}}
=\sum_{\sigma=\uparrow,\downarrow}\sum_{\langle i, j\rangle}t_{i,j}
\left(c_{i,\sigma}^{\dagger}c_{j,\sigma}+c_{j,\sigma}^{\dagger}c_{i,\sigma}\right)
+U\sum_{i}n_{i,\uparrow}n_{i,\downarrow}
+\mu\sum_{\sigma=\uparrow,\downarrow}\sum_{i}n_{i,\sigma}
\end{eqnarray}
with the on-site repulsion $U>0$.
We have chosen the sign of hopping terms $t_{i,j}>0$ in the Hubbard model (\ref{hubb_ham})
in order to emphasize the correspondence with the Heisenberg model (\ref{heis_ham}).
Note that the magnetic field $h$ in the Heisenberg model (\ref{heis_ham}) 
plays the same role as the chemical potential $\mu$ in the Hubbard model (\ref{hubb_ham}).
Below we consider the spin model (\ref{heis_ham}) in strong magnetic fields
around the saturation field $h_{\rm{sat}}$
and the Hubbard model (\ref{hubb_ham})
at values of the chemical potential around a characteristic value $\mu_0 $, see below.
Although both models represent different physics,
we will demonstrate that the mathematical consideration exhibits many common features.
The physical properties for the spin system for $h\approx h_{\rm{sat}}$ 
and for the electron system for $\mu\approx\mu_0$
will be governed exclusively by either localized magnon or electron states which exist due to lattice geometry.

We begin with the spin model (\ref{heis_ham}) \cite{prl,zhito:2004,dr:2004,ltp}.
The lowest excitations in strong magnetic fields $h> h_{\rm{sat}}$
are one-magnon states above the fully polarized ferromagnetic state $\vert{\rm{FM}}\rangle$.
The lower of the two branches of the one-magnon dispersion for the sawtooth chain becomes flat for $J_2=2J_1$.
For this case one can construct magnon states located in one of the valleys of the sawtooth chain
$\vert 2j\rangle=(s_{2j-1}^--2s_j^-+s_{2j+1}^-)\vert{\rm{FM}}\rangle$
with the energy $E_{\rm{FM}}-\epsilon_1-h(N/2-1)$,
$E_{\rm{FM}}$ is the energy of the state $\vert{\rm{FM}}\rangle$ of the spin system (\ref{heis_ham}),
$\epsilon_1=4J_1$.
Next, we consider the electron model (\ref{hubb_ham}) \cite{mielke,tasaki:1998,dhr}.
The lower one-electron band becomes completely flat for $t_2=\sqrt{2}t_1$.
The localized one-electron states can be written as
$\vert 2j,\sigma\rangle
=(c_{2j-1,\sigma}^{\dagger}-\sqrt{2}c_{2j,\sigma}^{\dagger}+c_{2j+1,\sigma}^{\dagger})\vert 0\rangle$
and their energy is $\varepsilon_1=-2t_1+\mu$.
Although the one-particle problem for both Hamiltonians is quite similar,
the many-particle problem will obviously be different.
For the Heisenberg system 
we deal with magnons which are hard-core bosons with nearest-neighbor repulsion
(see Eq.~(\ref{heis_ham})),
whereas for the Hubbard system
we deal with interacting spinful electrons which represent a two-component fermionic mixture with one-site repulsion
(see Eq.~(\ref{hubb_ham})).

\section{Localized states in the presence of interactions}

Since the localized states are located in a restricted area of the whole lattice,
it is clear
that many-particle states consisting of several isolated (no common sites) occupied valleys
are exact eigenstates of both Hamiltonians \cite{prl,zhito:2004,dr:2004,ltp,mielke,tasaki:1998,dhr}.
However, for the Hubbard model, by contrast to the Heisenberg model,
the localized states with the same spin polarization may also have common sites.
By direct computation one shows that,
e.g., $\vert 2j,\sigma\rangle\vert 2j+2,\sigma\rangle$
is indeed an eigenstate of the Hamiltonian (\ref{hubb_ham}) in the two-electron subspace.
Further localized two-electron states with one common site can be obtained
owing to SU(2) symmetry of the Hubbard model (\ref{hubb_ham})
by applying operators
$S^{-}=\sum_ic_{i,\downarrow}^{\dagger}c_{i,\uparrow}$
or
$S^{+}=\sum_ic_{i,\uparrow}^{\dagger}c_{i,\downarrow}$
on $\vert 2j,\sigma\rangle\vert 2j+2,\sigma\rangle$.
This example clearly shows a difference between magnons and electrons
conditioned by different particle statistics and interaction.
Finally,
we notice that a maximal number of localized magnons/electrons  
which can be put on the sawtooh-chain lattice
is $n_{\max}=N/4$ for magnons
but $n_{\max}=N/2$ (corresponding to quarter filling) for electrons;
here $N$ is the (even) number of sites of the sawtooth-chain lattice.

Since the localized states are the lowest-energy states in the one-particle subspace,
one may also expect that the states with $n$ isolated (independent) localized particles
are the lowest-energy states in the $n$-particle subspace
(for rigorous proofs see Ref.~\cite{schmidt}).
We can also confirm this by exact diagonalizations of finite systems \cite{prl,dr:2004,ltp,dhr}.
Moreover, numerics gives evidence
that for many lattices these ground states are separated from the higher-energy states by a gap.
Another important property of the localized states is their linear independence \cite{schmidt:2006}.
The energy of the localized $n$-particle states is
$E_{n}^{\rm{lm}}=E_{{\rm{FM}}}-hN/2+n(h-4J_1)$ 
for the Heisenberg model 
and  
$E_{n}^{\rm{le}}=n(\mu-2t_1)$ 
for the electron model. 
Obviously,
the localized-magnon states 
(localized-electron states) 
are degenerate at the saturation field $h=h_{\rm{sat}}=4J_1$ 
(at a characteristic value of the chemical potential $\mu=\mu_0=2t_1$).

Consider now the spin model in a strong magnetic field $h$ around the saturation field $h_{{\rm{sat}}}$.
Using the ensemble with fixed $(h,N)$ 
we find the following contribution of the localized-magnon states to the partition function
\begin{eqnarray}
\label{partit_ma}
Z(T,h,N)=\sum_{n=0}^{n_{\max}}g^{\rm{mag}}_{N}(n)
\exp\left(-\frac{E_{n}^{\rm{lm}}}{T}\right)
\propto\sum_{n=0}^{n_{\max}}g^{\rm{mag}}_{N}(n)\exp\left(\frac{h_{\rm{sat}}-h}{T}n\right).
\end{eqnarray}
Here $g^{\rm{mag}}_{N}(n)$ is the degeneracy of the states with $n$ independent localized magnons.
Thermodynamic quantities follow by the standard relations:
$F(T,h,N)=-T\ln Z(T,h,N)$,
$S(T,h,N)=-\partial F(T,h,N)/\partial T$ (entropy),
$C(T,h,N)=T\partial S(T,h,N)/\partial T$ (specific heat) etc.
Similarly,
we consider the electron model at a value of the chemical potential around $\mu_0$.
Using the (grand-canonical) ensemble with fixed $(\mu,N)$ 
we find the following contribution of the localized-electron states to the partition function
\begin{eqnarray}
\label{partit_el}
\Xi(T,\mu,N)=\sum_{n=0}^{n_{\max}}g^{\rm{el}}_{N}(n)
\exp\left(-\frac{E_{n}^{\rm{le}}}{T}\right)
=\sum_{n=0}^{n_{\max}}g^{\rm{el}}_{N}(n)\exp\left(\frac{\mu_0-\mu}{T}n\right).
\end{eqnarray}
Here $g^{\rm{el}}_{N}(n)$ is the degeneracy of states with $n$ independent localized electrons.
Thermodynamic quantities follow by the standard relations:
$\Omega(T,\mu,N)=-T\ln \Xi(T,\mu,N)$,
$S(T,\mu,N)=-\partial\Omega(T,\mu,N)/\partial T$ (entropy),
$\bar{n}(T,\mu,N)=\partial\Omega(T,\mu,N)/\partial \mu$ (average number of electrons) etc.
We note that the specific heat $C(T,n,N)$ at constant $n$ equals zero for $n=1,\ldots,N/2$.

The central problem now is the calculation of the degeneracy $g_N(n)$ 
for localized magnon and electron states.
This can be done using a mapping of localized states onto spatial configurations of hard dimers on a simple chain.
For the spin system it can be shown that
$g^{\rm{mag}}_N(n)=Z(n,N/2)$,
$n=0,1,\ldots,N/4$,
where $Z(n,{\cal{N}})$ is the number of spatial configurations of $n$ hard dimers on a chain of ${\cal{N}}$ sites
\cite{zhito:2004,dr:2004,ltp}.
For the electron system it can be shown that
$g^{\rm{el}}_N(n)=Z(n,N)$,
$n=0,1,\ldots,N/2-1$,
but
$g^{\rm{el}}_N(N/2)=N/2+1$
\cite{dhr}.
Substituting $g^{\rm{mag}}_N(n)$ and $g^{\rm{el}}_N(n)$ into Eqs.~(\ref{partit_ma}) and (\ref{partit_el})
we obtain a grand-canonical partition function of one-dimensional hard dimers,
which can be calculated using the transfer-matrix method.
As a result,
we obtain the low-temperature thermodynamics for both models.

\begin{figure}
\begin{center}
\includegraphics[width=17.7pc]{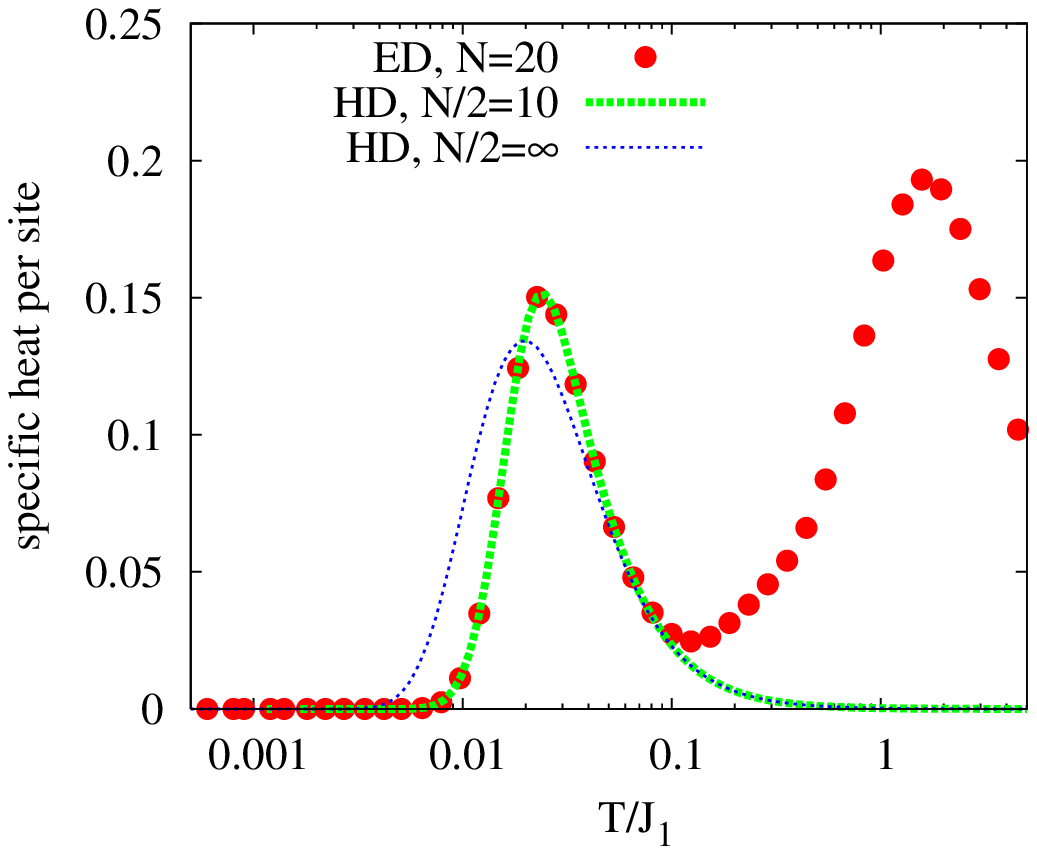}
\hspace{1pc}
\includegraphics[width=17.7pc]{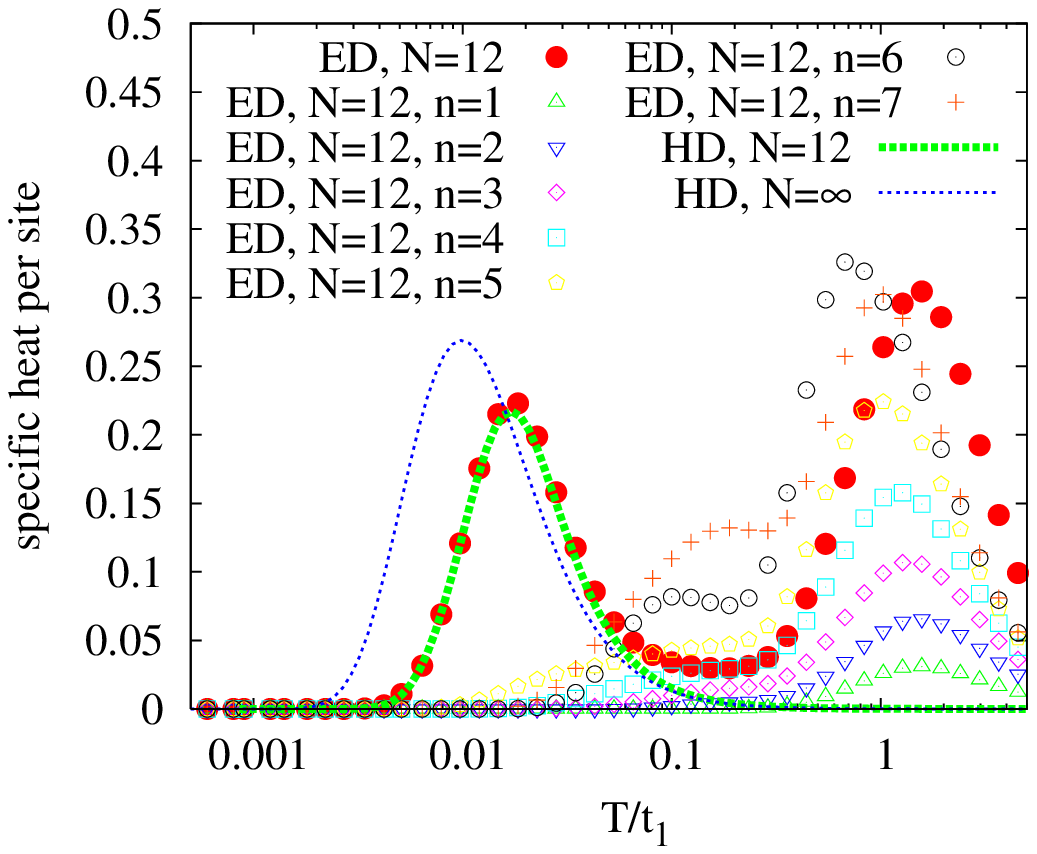}
\caption{\label{fig2}
Specific heat
for the spin model (\ref{heis_ham}) ($J_1=1$, $J_2=2$)
at $h=0.98h_{\rm{sat}} = 3.92 J_1$ (left)
and the electron model (\ref{hubb_ham}) ($t_1=1$, $t_2=\sqrt{2}$)
around $\mu_0=2t_1$ (right).
Left:
exact diagonalization data for $N=20$ (filled circles);
hard-dimer data for $N/2=10$ (thick curve) and $N/2\to\infty$ (thin curve).
Right:
exact diagonalization data for $U\to\infty$, $N=12$
for $C(T,n,N)/N$,  $n=1,2,3,4,5,6,7$ electrons
(up-triangles, down-triangles, diamonds, squares, pentagons, circles, and crosses, respectively)
and
$C(T,\mu=0.98\mu_0,N)/N$ (filled circles);
hard-dimer data for $N=12$ (thick curve) and $N\to\infty$ (thin curve)
for $C(T,\mu=0.98\mu_0,N)/N$.}
\end{center}
\end{figure}

In Fig.~\ref{fig2} we illustrate the localized-state predictions (\ref{partit_ma}) and (\ref{partit_el})
for the temperature dependence of the specific heat.
The low-temperature maximum in Fig.~\ref{fig2} is due to the manifold of localized states.
Obviously,
the localized-state picture excellently reproduces exact diagonalization data for finite systems
in the low-temperature regime 
for small deviation from the values $h_{\rm{sat}}$ or $\mu_0$.
Moreover, we note that localized-state thermodynamics implies 
an enhanced magnetocaloric effect \cite{cooling,cooling2}.

\section{Summary}

To summarize,
we have studied the localized-state effects
for two strongly interacting models
(antiferromagnetically interacting Heisenberg spins and standard Hubbard electrons)
on the sawtooth-chain lattice which supports a flat one-particle band.
Under certain conditions
(values of external magnetic field or chemical potential)
the localized states in both models govern the low-temperature thermodynamics.

Several remarks are in order here.
The large-$U$ limit of the Hubbard Hamiltonian (\ref{hubb_ham}) yields the $t$--$J$ model.
Therefore, it is not astonishing
that the localized-electron states are also eigenstates of the $t$--$J$ Hamiltonian
and they are the ground states for $n=1,\ldots,N/2$ electrons
for small exchange couplings $J_{i,j}$ \cite{andreas}.
Some prominent peculiarities of low-temperature thermodynamic quantities conditioned by localized states
remain stable to small deviations from the ideal lattice geometry,
increasing chances to observe the examined properties in solid-state systems \cite{dr:2004}.
Although we focus here on one representative example,
the sawtooth-chain lattice,
the elaborated scheme can also be applied to spin and electron models on other lattices \cite{ltp,other}.

\ack
We thank J J\c{e}drzejewski for discussions.
O~D acknowledges the kind hospitality
of Magdeburg University (2008) 
and
of the MPIPKS (Dresden) during the seminar
Unconventional Phases and Phase Transitions in Strongly Correlated Electron Systems (2008).
A~H is supported by the DFG through a Heisenberg fellowship (Project HO~2325/4-1).

\end{document}